# A novel concept for a fully digital particle detector

G. Casse,[a,b] N. Massari,[a] M. Franks[a,b] and L. Parmesan[a]

[a] *Fondazione Bruno Kessler (FBK),*
 *Via Sommarive 18, 38123, Trento, Italy*
[b] *University of Liverpool, Department of Physics,*
 *O. Lodge Lab., Oxford Street, L69 7ZE, Liverpool , UK.*
 *E-mail*: gcasse@liv.ac.uk

ABSTRACT: Silicon sensors are the most diffuse position sensitive device in particle physics experiments and in countless applications in science and technology. They had a spectacular progress in performance over almost 40 years since their first introduction, but their evolution is now slowing down. The position resolution for single particle hits is larger than a few microns in the most advanced sensors. This value was reached already over 30 years ago [1]. The minimum ionising path length a sensor can detect is several tens of microns. There are fundamental reasons why these limits will not be substantially improved by further refinements of the current technology. This makes silicon sensors unsuitable to applications where the physics signature is the short path of a recoiling atom and constrains the layout of physics experiments where they represent by far the best option like high energy physics collider experiments. In perspective, the availability of sensors with sub-micron spatial resolution, in the order of a few tens of nanometres, would be a disruptive change for the sensor technology with a foreseeable huge impact on experiment layout and various applications of these devices. For providing such a leap in resolution, we propose a novel design based on a purely digital circuit. This disruptive concept potentially enables pixel sizes much smaller than $1\mu m^2$ and a number of advantages in terms of power consumption, readout speed and reduced thickness (for low mass sensors).

# Contents

## 1. Introduction

Silicon detectors consist of three functional parts: the sensing cell, the analogue amplifier and the on-chip digital electronics. The sensing elements are arrays of reverse biased pn junctions (diodes). They collect the charge released by ionising particles in their volume. Typical diode dimensions in micro-strip arrays are 50 µm to 500 µm in width and 1 cm to 10 cm in length [2], while in pixels they range from 30x30 µm$^2$ to 500x500 µm$^2$ [2, 3]. Typical thicknesses go from 150 to 300 µm. The small signal generated by the ionised charge is amplified by analogue circuits connected to each diode. The signal is then digitised and processed for final readout and storage. There is important digital activity locally on-chip, dedicated to charge cluster (hits distributed over two or more channels) finding, buffering, trigger algorithms, data compression, time stamping, data streaming etc. The three functional parts of the sensor can be implemented on separate wafers or on the same substrate.

The analogue and digital electronics are realised in CMOS technology, while the sensing part can be produced in standard or custom technologies, available in niche detector foundries. The quest for increasingly better position resolution and fine granularity coming from physics experiments and applications has motivated the reduction of the pixel size. Current state-of-the-art sensors have typical pixel sizes of 50x50 µm$^2$, in the case of hybrid pixels [4], and 28x28 µm$^2$ in monolithic sensors [5]. The accelerated evolution of the microelectronics industry with decreasing feature size S (defining the minimum channel length of a MOS transistor) has enabled these relatively small pixel dimensions. As the number of transistors per unit area typically scales with $1/S^2$, this permitted an enormous increase in the processing power of the digital circuitry with equivalent surface. On the other hand, the analogue part cannot be miniaturised at the same level, as small transistors introduce noise and mismatched performances when shrinking the transistor size. Moreover, the decreasing voltage supply of modern technologies may affect the dynamic range and accuracy of the circuit especially when area constraints drive the design. In conclusion, the analogue transistors are over-dimensioned by design, being often orders of magnitudes larger than the minimal size offered by the technology. For these reasons, the 65 nm is currently the smallest useable technology node [6], while the most advanced CMOS technology now available for digital electronics is the 3 nm node. The analogue circuit is the actual limiting factor for further significant reduction of the pixel size and consequent improvements of resolution and sensitivity of the sensors.

## 2. A paradigm change: digital only sensors for deep sub-micron tracking capabilities

To enable a spatial resolution in the range of tens of nm, corresponding to an improvement of almost two orders of magnitude, a possible, if not the only, route is to abandon analogue amplification for sensing the small signals induced by ionising radiation. This removes the need for oversized transistors, so that sensor circuits can exploit the reduction of the feature size in CMOS nodes smaller than the 65nm one. The circuits would be substantially simplified, as the number of transistors needed for each sensing channel could be reduced to a few, yielding extremely small pixels. In this view, the design effort will focus on moving the analogue to digital conversion point as close as possible to the sensing part. The key point of this concept is to conceive a circuit capable of efficiently detecting the hit position from its ionisation trail in silicon and convert it directly to a digital signal (a single bit) for subsequent processing steps. We introduce here a novel sensor concept inspired by the memory cell technology to devise an entirely different detection mechanism.

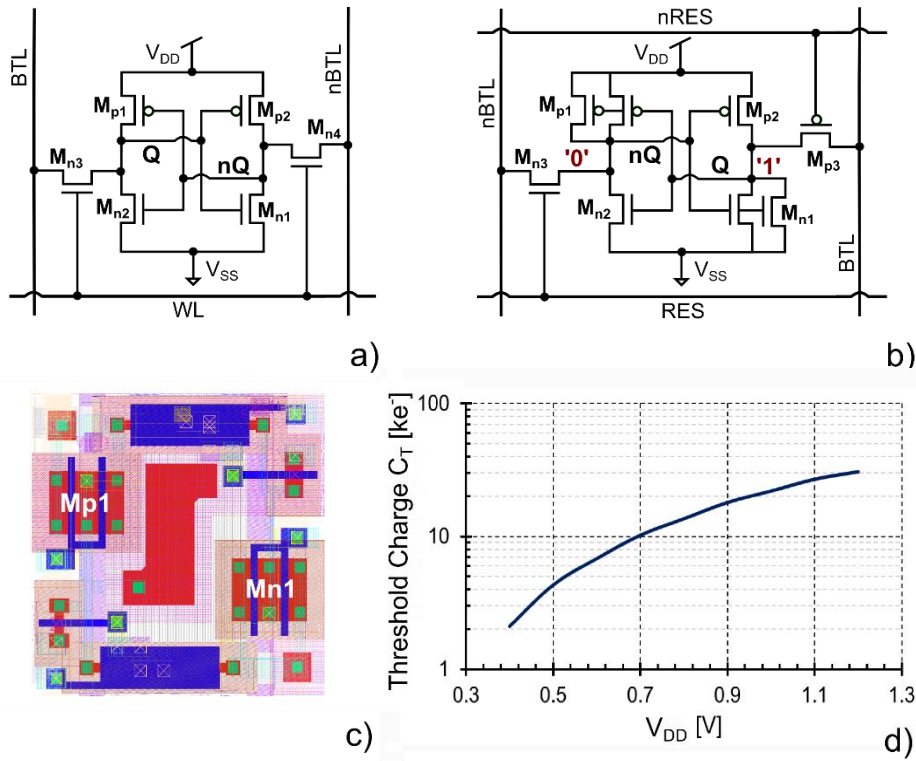

**Figure 1.** a) Schematics of a standard SRAM; b) Modified SRAM for improved sensitivity to radiation; c) layout of a single cell in UMC 65nm; d) simulations of the minimum charge ($C_T$) needed for toggling the cell as a function of the bias voltage $V_{DD}$.

### 2.1 The digital sensor concept

Memory cells are digital circuits with two selectable states (0 and 1) used to store a single bit of information. The typical layout of the widely used Static Random Access Memory (SRAM) is shown in Fig. 1a. Four central transistors ($M_{n1}$, $M_{p1}$, $M_{n2}$, $M_{p2}$) form two cross-coupled inverters. The operations are performed by setting the Write Line (WL) that controls the gate of transistors

Mn3 and Mn4. The bit lines BTL (nBTL) line is driven high (low) by the inverter circuit. For writing, a logic '1' or '0' is applied to the bit lines, the WL is asserted and the set value is stored. Memories are subjected to occasional, undesired state changes induced by crossing ionising particles. These events are called Single or Multiple Event Upset (SEU or MEU) depending on the number of digital nodes affected [7]. They cause problems (soft errors) in electronics with high transistor density. Digital circuits are designed to be robust against these, by implementing temporal or spatial redundancy, dummy structures or other mechanisms for correcting their occurrence. Thanks to these measures, SEUs occur at very low rates in modern electronics. Instead, we want a digital device with very high probability of a state change induced by crossing ionising particles.

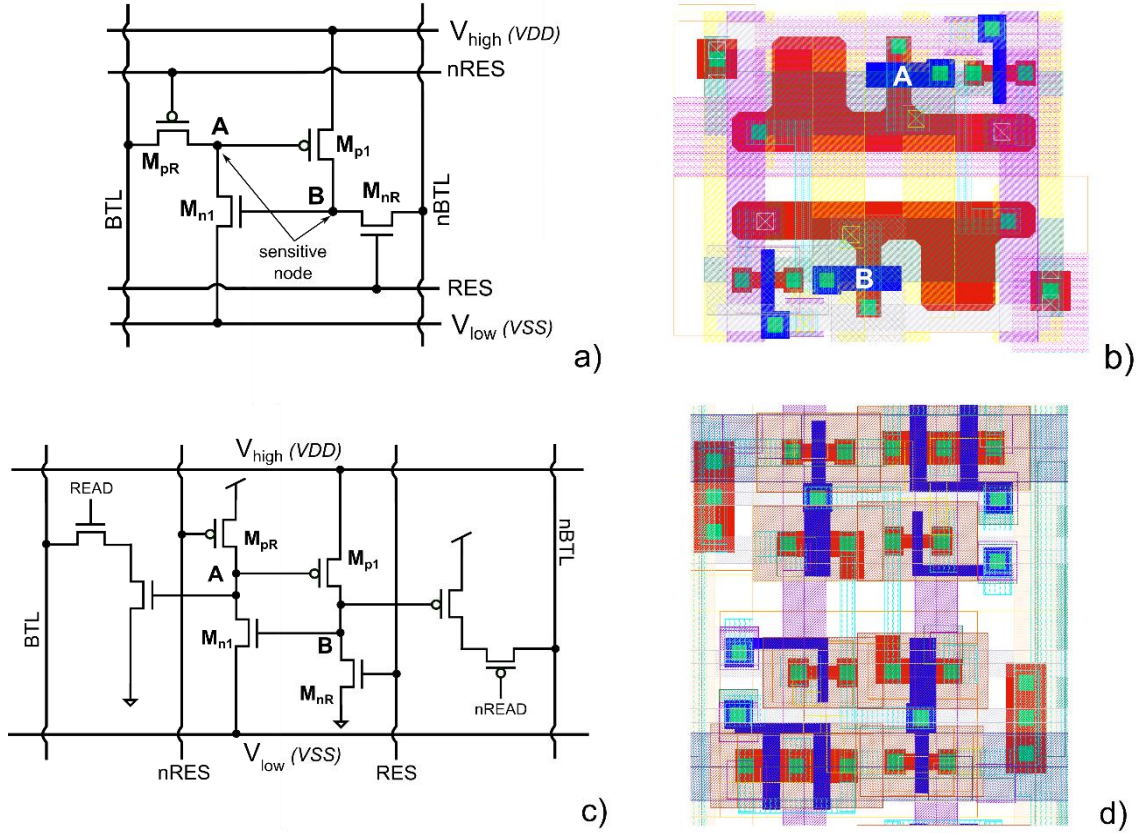

**Figure 2.** Schematics of implemented digital sensor (Dsens) cells: a) circuit level schematic of the improved Dsen1 cell; b) cell layout of the improved Dsen1 cell using UMC 65 nm; c) circuit level schematic of the basic Dsen2 cell ; d) cell layout of the basic Dsen2 cell using UMC 65 nm.

The performance parameters to assess the detection properties are the Threshold Charge ($C_T$) that is the minimum charge required for toggling the circuit and the subsequent Switching Probability (SP) of a state flip when a particle crosses the detector. Another important parameter is the dark rate ($D_R$), defined as the number of random, unprovoked state changes. At design stage, the estimate of $C_T$ and SP for given ionising events is not straightforward as they depend on the circuit schematics, the CMOS feature size S, the substrate resistivity, the dimensions and parasitic characteristics of the circuit (like the parasitic capacitance of the sensing node) and the operation conditions (bit line and world line voltages, gate pre-charging). Several layouts are possible in various CMOS nodes, introducing multiple variables in the investigation of the new concept.

We have simulated a few innovative topologies to reduce $C_T$ for maximum efficiency and sensitivity. The layout in Fig. 1b and 1c is an example of a modified SRAM with improved sensitivity to ionising events. The operation differs from an SRAM because it is initialized to a known reset condition waiting for an ionising particle. Transistors $M_{n3}$ and $M_{p3}$ of Fig. 1 set the internal sensitive nodes nQ and Q to '0' and '1' respectively. When a particle hits the cell, a voltage change to the internal nodes is generated by the ionised charge causing the memory to flip. The charge collected at the sensitive nodes needs to be high enough not only to switch ON transistor $M_{n1}$ ($M_{p1}$) but also to create an overdrive voltage to switch the $M_{p2}$ and $M_{n2}$ transistors that are, in the initial state, both ON. To facilitate this, $M_{n1}$ and $M_{p1}$ are designed with higher transconductance to favour the switch. The overdrive voltage depends on the power supply voltage $V_{DD}$. Our simulations show (Fig. 1d) a steep dependence of $C_T$ on the voltage supply reaching a minimum of about 2 ke$^-$ for low values (0.4 V) of $V_{DD}$.

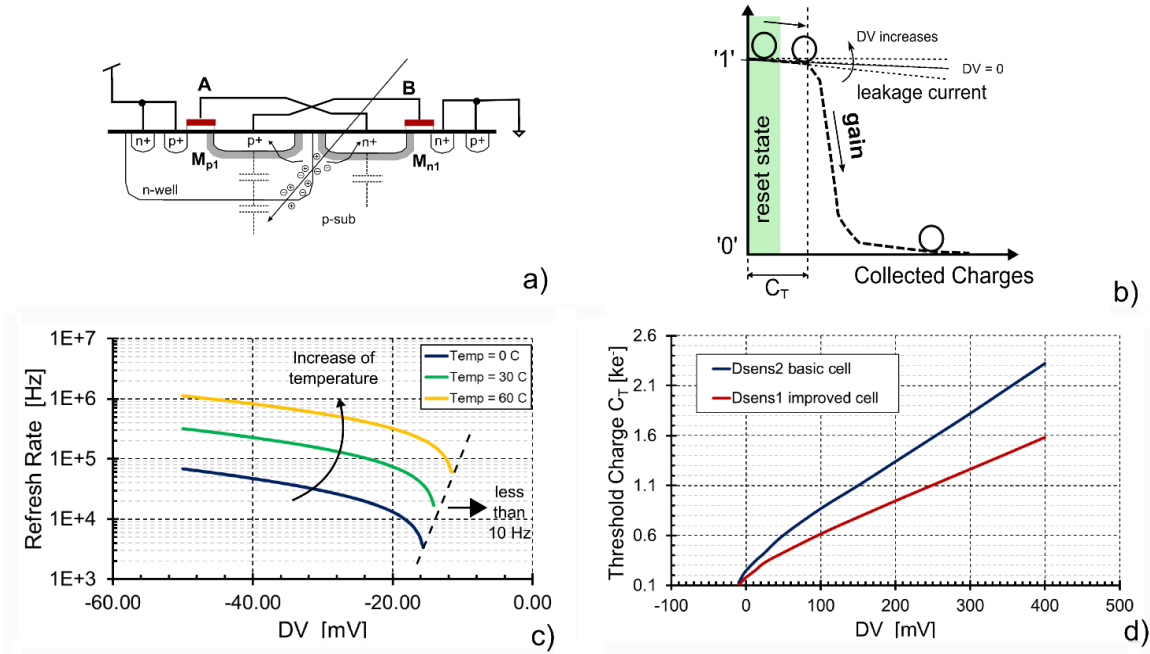

**Figure 3.** a) Cross section of the Dsen1 cell with a crossing particle; b) description of the working principle of DSen1; c) refresh rate required to suppress false hits (dark rate) induced by leakage current at different temperatures; d) simulation of the Threshold Charge (CT) at node A needed to flip the cell state for two different pixel designs (Fig. 4 a) and b)).

Better sensitivity can be achieved with designs that move away from the SRAM standard layout. We envisaged a novel design (Digital Sensor, DSen1 - Fig. 2) consisting of only four transistors, thereby considerably reducing the footprint. Differently from SRAM, this cell operates as a volatile memory that needs periodic refreshing. Transistors $M_{nR}$ and $M_{pR}$, driven by the complementary signals RES and nRES respectively, set the initial state forcing the node A (B) to the predefined value $V_{DD}$ ($V_{SS}$). After reset, the sensitive nodes A and B are left floating. They are connected to the p+/n-well junction of the drain of transistor $M_{p1}$ and to the n+/p-sub junction of transistor $M_{n1}$ (Fig. 2b-c). To improve SP, the area of these junctions can be augmented for increased charge collection. This also increases the parasitic capacitance and can induce a slight worsening of SP. TCAD simulations have been used to select the best geometry.

The cell operates as a bistable circuit, as explained in Fig. 3. In this representation the circle (Fig.3b) is the status of the cell, initially placed at the reset state ('1'). When the charge generated by a crossing ionising particle is collected at node A (B) the circle is shifted to the right proportionally to the amount of charge seen on the nodes. If the charge is equal or higher than $C_T$, the transistor $M_{p1}$ ($M_{n1}$) is switched on and a positive feedback is triggered to irreversibly flip the memory (state '0'). As shown in Fig. 3b, the reset condition is not a flat plateau, but has a small slope due to leakage current. This would cause false events if the observation time were too long, requiring periodic reset. A careful simulation of the cell using different transistor sizes and types available in the chosen technology and offering low-leakage and low-power features were done to minimize this effect. Besides, the leakage can be controlled by means of a small voltage difference (DV) added or subtracted to the reference voltages, making $V_{low} = V_{ss}+DV$ and $V_{high} = V_{DD}-DV$. A positive value of DV forces transistors $M_{n1}$ and $M_{p1}$ to a strong OFF state after reset, reducing the leakage current, while a negative DV reduces the value of $C_T$. Scanning DV allows for finding the best trade-off between low $C_T$ and workable refresh rate. Figure 3c shows the refresh rate as a function of DV needed to avoid false triggers with an optimized design cell (in 65nm UMC) operated at different temperatures. We fixed a target refresh rate ≤ 10 Hz for our operations, so forcing the choice of DV > -10 mV or positive. This satisfies our criterion for operations at temperatures ranging from 0 to 60 °C (with visible advantage at low temperatures). Figure 3c shows the minimum charge needed for switching the cell for two different layout implementations as a function of the admissible range of DV. Simulations predict a low $C_T$ value in the range of 100 e$^-$.

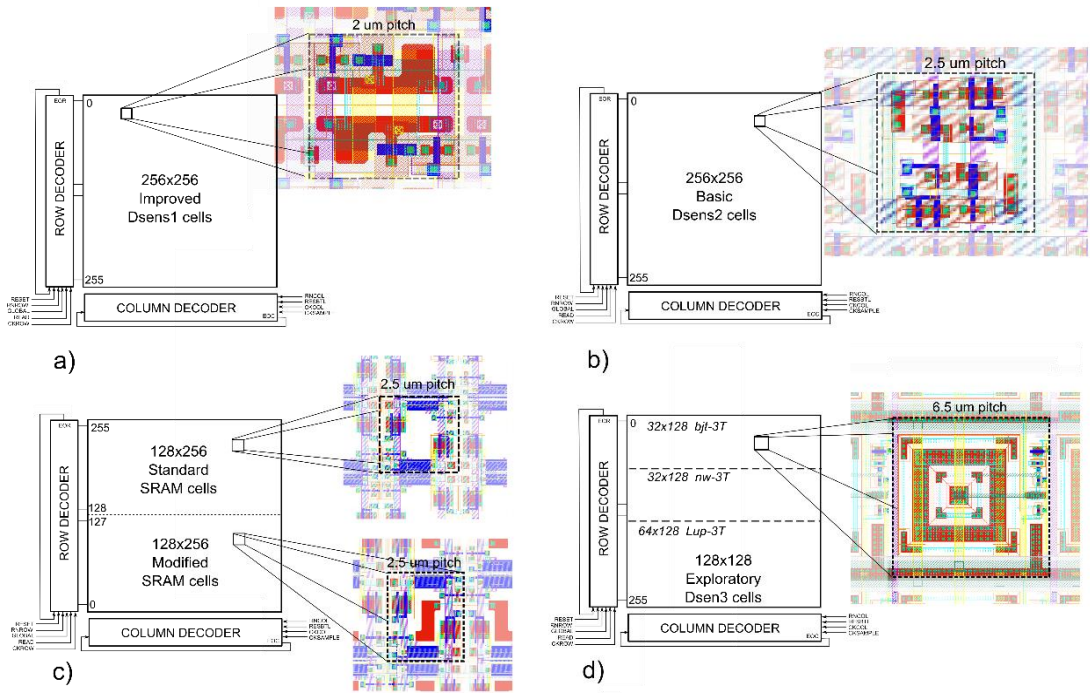

**Figure 4.** Prototype detector with four different arrays: a) 256x256 improved (extended collected nnodes) 4T type DSen1; b) 256x256 basic 8T DSen2 c) 128x128 standard and 128x128 modified SRAM; d) 128x128 exploratory Dsen3.

**2.2 The digital sensor prototype**

These TCAD simulations have been performed using our best knowledge of the 65 nm UMC CMOS. Although the incomplete availability of all processing details of the technology introduces uncertainties, the outcome of the simulations is very encouraging regarding the feasibility of this fully digital particle sensor and stimulated the design of a few explorative layouts for an MPW submission. The test chip (Fig. 4) consists of four arrays. Fig. 4a) and b) show the improved and basic digital sensor layouts, DSen1 and DSen2. The basic implementation uses an 8-transistor

– 5 –

design and has a footprint of 2.5x2.5 μm$^2$. The improved layout has only four transistors with a cell size of 2x2 μm$^2$ and an extended collecting junction size to increase the effective sensitive area. It must be stressed that the parasitic capacitance of the sensitive node in the basic pixel is lower than in the improved one. The reference array contains two implementations of an SRAM with a cell footprint of 2.5x2.5 μm$^2$ (Fig. 4c). The remaining array includes three different exploratory cell topologies. The third 128x128 array (DSen3, Fig.4 d) has a larger pixel area of 6.5x6.5 μm$^2$ and it is designed to exploit the presence of parasitic vertical and lateral BJT transistors for implementing the sensitive node to particle hits.

All arrays include four test pixels with a calibrated injection circuit for characterizing the response to a known charge applied to the sensitive nodes for the estimation of the value of $C_T$ needed to toggle the cell (Fig. 5). The arrays are read out by row and column decoders that address all pixels in a raster scan manner.

This prototype sensor has been submitted for an MPW production run with the UMC [8] 65 nm low-leakage technology accessed through Europractice [9]. No customised solutions were possible to improve the detector performance such as increased substrate resistivity to enhance charge collection at the sensitive nodes, optimised design for introducing an electrical drift field beneath the collecting electrodes etc. The layout has been prepared according to the foundry design rules and available substrate resistivity (nominally 7 – 17 Ω· cm). This implies that the charge seen at the sensitive transistor nodes is the fraction of carriers ionised in a small volume underneath the gate with negligible contribution from charges diffusing from the substrate. Nonetheless, we used this sub-optimal solution to test and validate our ideas and initial simulations.

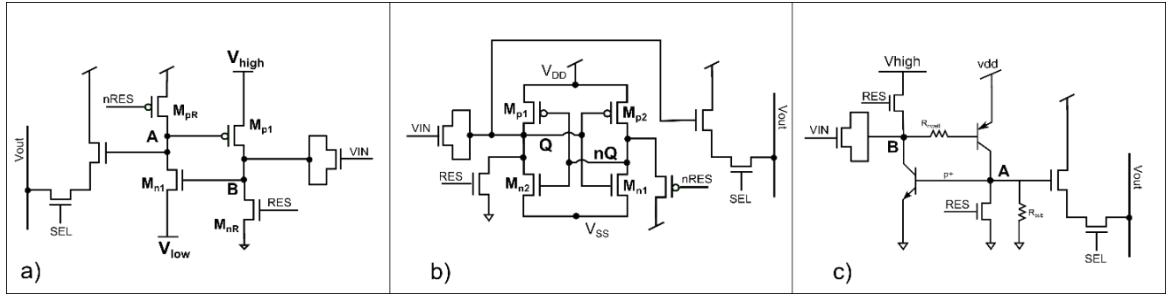

**Figure 5.** a) Schematic of the testing circuit of the Dsen1 and Dsen2 structures; b) Schematic of the testing circuit of the standard SRAM cell; c) Schematic of the testing circuit of Dsen3 with parasitic bjt.

## 3. Experimental results

The first functional test performed on a few sensors from the UMC submission was the estimate of $C_T$ using the charge injection circuit. Figure 6 shows the typical output of one set of four cells responding to the same charge injection. For a fixed value of DV (100 mV). The switching probability as a function of the injected charge has the expected S-curve shape and we define the minimum charge $C_T$ for toggling each pixel as the value where the probability crosses the 50% level. The observed spread of $C_T$ is mainly due to transistor mismatch, and it is expected in real devices due to leakage current fluctuations. For this reason, we operated the sensors with a minimum DV of 70 mV. In the current layout DV is common across all pixels of a given array, while in future sensors it could be used for individual trimming of pixels though a calibration procedure.

Figure 7 shows $C_T$, calculated as the average of the four test pixels as a function of DV for DSen1, compared with $C_T$ extracted from a custom SRAM as a function of $V_{DD}$. In both structures, a decrease of DV and $V_{DD}$ determines an increase in the cell sensitivity. The minimum $C_T$ for the DSen1 cell is below 1 ke$^-$, while for the SRAM is about ten times larger, as already predicted by simulations, (Fig. 1d)), showing the large improvement achieved with the DSen layout. A $C_T$ under 1ke$^-$ is already a good value for many detector applications.

As mentioned before, the leakage current of the circuit can induce spurious events when the observation time is sufficiently long. Figure 8 shows the average refresh rate needed to suppress dark events in all four test pixels. The intensity of the leakage current can be reduced increasing the value of DV. The choice of DV needs to be optimised in consideration of the competing needs for lowering $C_T$ while keeping a comfortable refresh rate. Integration times exceeding 1 s are achieved with a DV of approximately 100 mV. The dark rate is also reducing by lowering the temperature, giving a second handle for controlling this rate to ensure very low dark-count also in slow frame-rate operations.

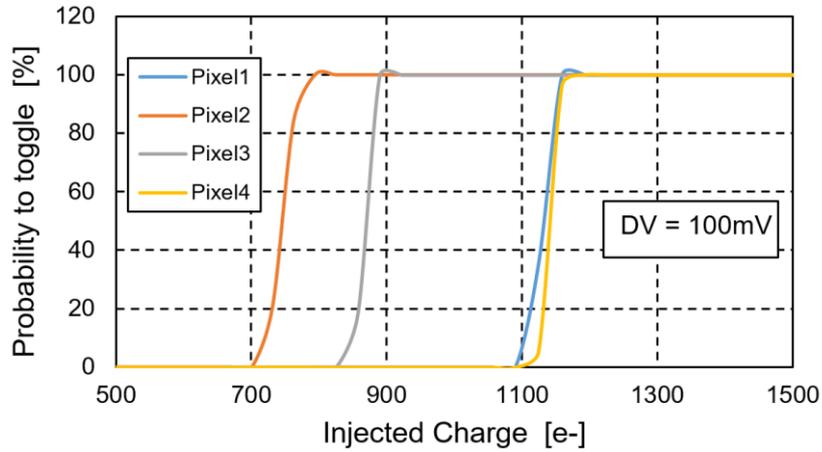

**Figure 6.** Typical switching probability of four test pixels as a function of the injected charge, with DV = 100 mV. The value of charge at which the curves cross the 50% probability is taken as $C_T$.

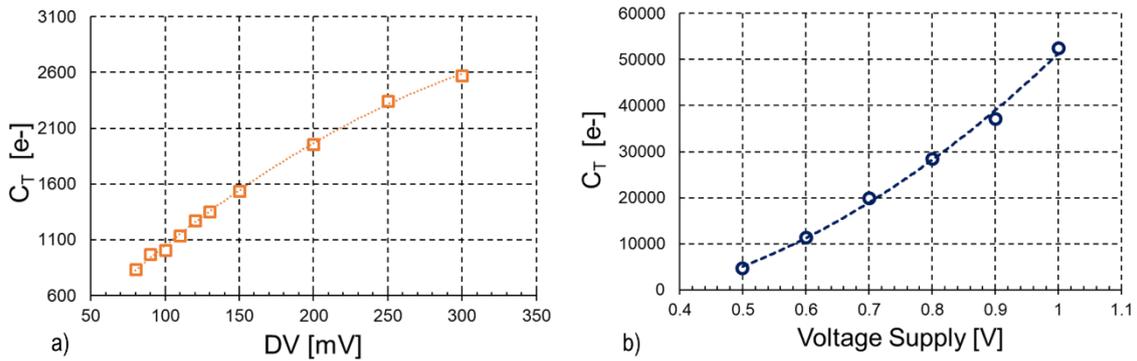

**Figure 7.** (a) Threshold charge $C_T$ as a function of DV for Dsen1 and (b) $C_T$ as a function of the voltage supply ($V_{DD}$) for a custom SRAM. The measurement shows an improvement of roughly one order of magnitude going from a custom SRAM to the Dsen1 cell.

### 3.1 Response to pulsed laser

The Dsen1 sensors have been tested using a pulsed blue laser (480 nm) with the setup shown in Fig. 9a). The sensor was running at 1.8 kfps and the laser was triggered at every frame to produce a burst of pulses. Figure 9b) shows the response of the array to a burst of 10 pulses applied at every integration time with the laser focused on the active area. A single frame acquired by the system is the result of the sum of 32 sequential binary images. In this way, the maximum value per pixel in the histogram is 32. The laser spot includes about 15x15 pixels (see the blow-up of Figure 9b)), for an equivalent spot size of 30x30 µm$^2$. The carrier injection conditions are difficult to estimate. The laser light spot is circular with a Gaussian intensity distribution from the centre. For testing the detector response to lower intensity signals, we measured the device output with two and even one laser shot per frame. Figure 9 shows the detector response to the various injection conditions, exhibiting a reduced size of the measured beam spot when only the central pixels are triggered by the higher intensity light at the centre of the distribution.

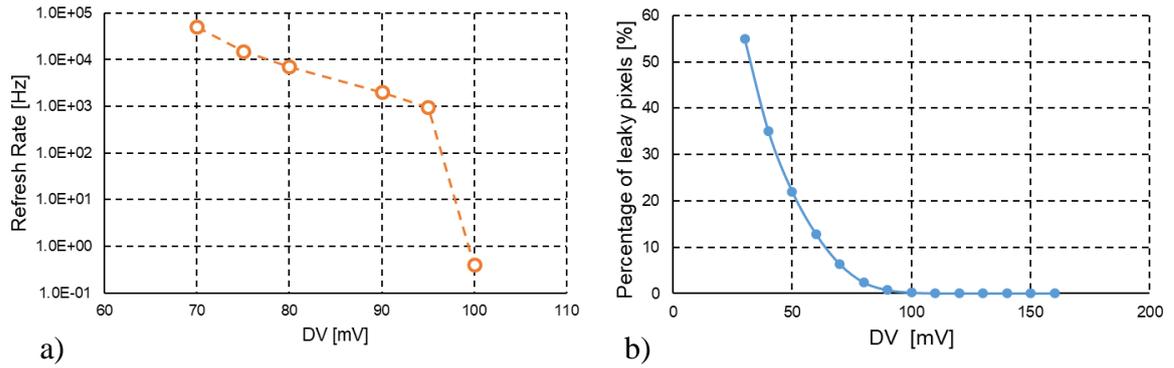

**Figure 8.** a) Refresh rate for dark rate suppression as a function of DV for the DSen1 layout; b) percentage of firing pixels in no-illumination conditions as a function of DV.

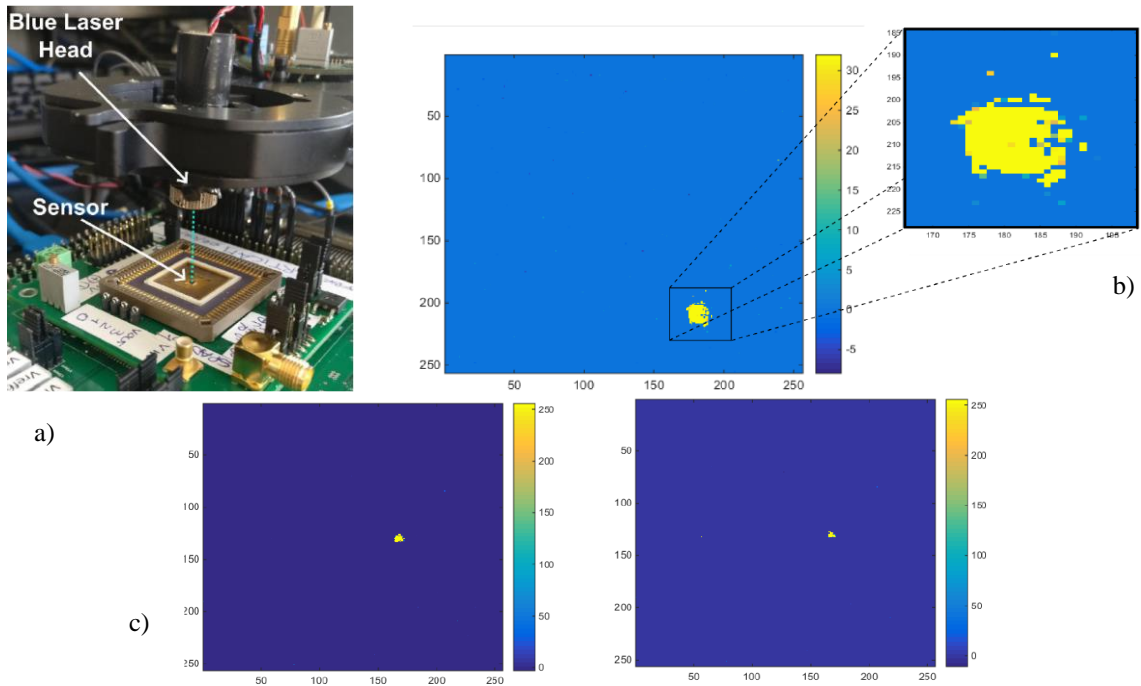

**Figure 9.** a) Laser experimental setup with DSen1. b) Response to laser with 10 pulses for each integration time (>DV ~ 100mV). c) Response of the sensor to two and a single laser pulse (DV ~ 100mV).

These results have been compared with the standard array that include pixel designed as regular SRAM cells. The array does not respond to the laser light injection, and tests with the charge injection circuits show a $C_T$ exceeding 6000e$^-$ even with a reduced voltage supply (0.5 V instead of
– 8 –

the 1.8 V typical for the technology). This underlines the importance of the layout for achieving sensitivity to ionising particles, while the standard memory circuits have a very small switching probability.

### 3.2 Response to alpha particles

The DSen1 array was further exposed to alpha particles from a $^{241}$Am radioactive source positioned at a fixed distance from the sensor. To prove the response to particles and show at the same time the insensitivity to dark noise, the sensor was partially covered with a metal absorbing sheet positioned between the source and the sensor and covering about half of the active area. The sensor response was studied in terms of cumulated hit counts over 256 kframes (Fig. 10 a)).

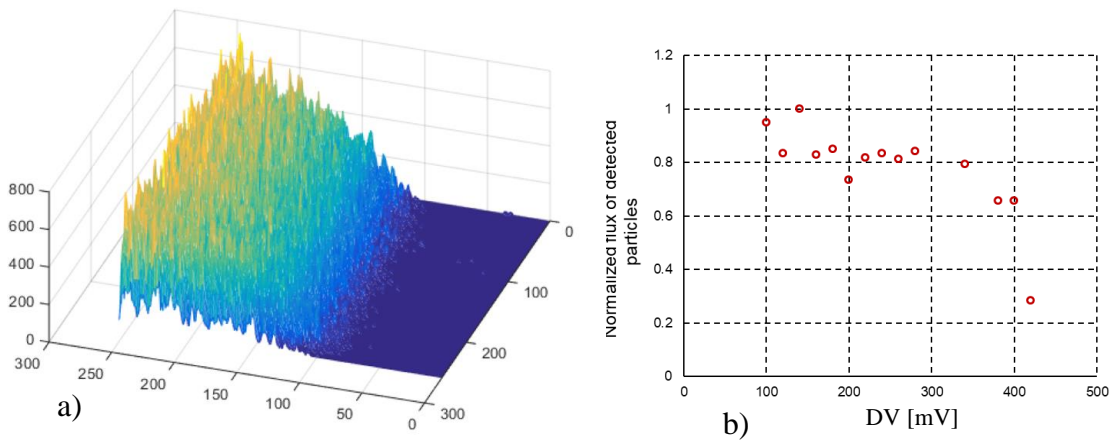

**Figure 10.** Response of DSen1 to alpha particles: a) A metallic shield is interposed between the source and the detector in order to cover about half of the surface to the radiation. b) variation of the detected flux of particles as a function of DV.

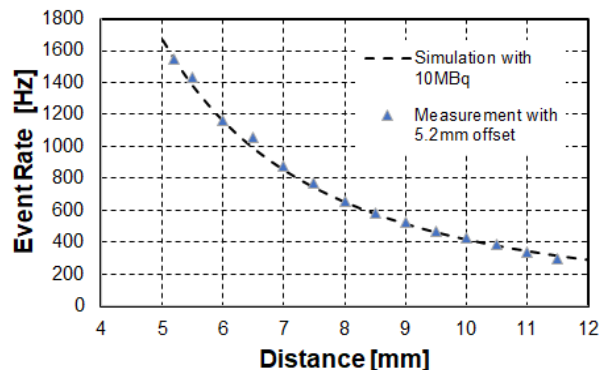

**Figure 11.** Alpha particle flux over the surface of a reference detector at various distances from the source. The dotted line is calculated assuming the nominal activity of the source (10MBq) and the points are measurements with a fast diode.

The portion of the detector not exposed to radiation is clearly visible and practically no spurious hits are recorded over the observation time. The smoothed edge of the signal is due to the partial coverage of the metal shield for alpha particles hitting the absorber at an angle, close to the edge. The majority of events are single hits with only about 1% of double hits rate. This indicates low or no charge sharing among neighbouring pixel supporting the interpretation that the sensitivity region is limited to the extension of the sensitive node area. Figure 10 b) shows the effect of varying DV on sensor response efficiency, which is decreasing with increasing DV. Operations with DV between 100 to 200 mV guarantee a very efficient dark rate suppression while keeping good efficiency. A sharp cut in detection efficiency is seen for DV values above 300 mV. We assessed the detector response with respect to the alpha particle rate. We estimated the alpha flux over the detector surface at different distances considering the nominal activity of the source (10 MBq) and verified it with a fast silicon diode as reference. The rate curve is shown in Fig. 11. The reference sensor, when positioned at 6.5 mm from the source, shows an alpha rate of about 1050Hz. On the other hand, Dsen1 measures a rate of about 100 Hz, a factor of 10 lower at the same distance.

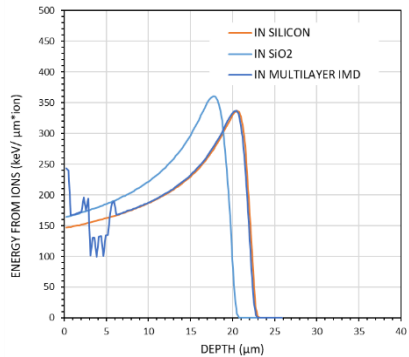

**Figure 12.** Ionising energy loss for $^{141}$Am alpha particles located at 6.5 mm from the source. The SRIM simulations show loss in SiO$_2$, pure Si and the multilayer structure describing the real digital sensor.

Figure 12 shows the SRIM [10] simulations of the ionization rate for alpha particles from a $^{141}$Am source located at 6.5 μm from the surface of the sensor. The three curves correspond to the energy loss in SiO$_2$, pure silicon and the multilayer representing the prototype device. This last curve is in fact really close to pure silicon. The multilayer curve includes the passivation layer on top and the various metals and dielectrics of the 65 nm UMC technology. The sensitive node implant sits at about 8 μm depth, where the ionization rate is approximately 175 keV/μm for an ionised charge of ~ 67 ke$^-$/μm. The Bragg peak is located at a depth of 21 μm with an energy loss of 335 keV/ μm (for an ionised charge of 128 ke$^-$/ μm).

This high charge density would cause a prompt switch of the cell if efficiently collected at the node. The reduced $S_P$ measured with alpha indicates that the charge reaching the sensitive node is a small fraction of the ionized one, likely released in the junction area of the node itself. Improving the node efficiency is a key point for using digital sensors for particle sensing applications.

In fact, we attribute the reduced rate measured with DSen1 to charge collection inefficiency due to the absence of a drift field to move charges towards the collecting electrodes and to the high recombination rate in low resistivity undepleted silicon. Under these hypotheses, we can consider as sensitive only the junction area with respect to the whole pixel one, with practically no depleted depth underneath. Also, we assume that one of the two sensitive nodes (the p-mos, which would be sensitive to positive charge collection) is less efficient due to the reduced hole mobility that makes them more subject to recombination. The ratio of the sensitive junction areas (n and p-mos) to the pixel surface is 30.9%. If only n-mos is collecting ionised charge the geometrical efficiency is about 15%, which is closer to the recorded efficiency of DSen1 to alpha particles.

These first results with laser and alpha particle illumination prove that the idea at the core of a fully digital pixel sensor for ionising radiation is feasible, with the sensor responding to both. The next steps will be the increase of the collection volume to significantly improve the response efficiency of purely digital sensors to impinging radiation.

## 4. Summary and outlook

This paper presents for the first time the concept of a fully digital sensor for ionising radiation. The purpose was to validate the concept showing that the digital response to ionising event is triggered by the relatively small charge released by crossing particles and photons. The first sensors have been produced with an MPW submission using standard UMC 65 nm technology. The concept has been validated showing response to 410 nm laser light and alpha particles from a $^{241}$Am source. The design does not optimize the charge collection efficiency at sensitive nodes, due to the limited customisation enforced by the MPW submission with the chosen technology. Improvements and further studies with future designs will be for investigating more advanced CMOS technologies (< 40 nm) and defining smart sensing structures able to collect charges deeply in the silicon substrate using, for example, a drift bias voltage to enhance the collection speed and effective volume.

The potential for a position sensitive radiation detector that does not make use of analogue amplification is enormous. Power consumption is at the minimum achievable while the readout speed can be maximized for a given pixel density in consideration of the binary output of every cell. The pixel size can be as small as hundreds of squared nanometres and the use of advanced CMOS nodes can be not only possible, but likely advantageous thanks to the expected significant reduction of $C_T$ in smaller feature size technologies. Further simulation studies of new layouts and solutions for improving charge collection to enable better efficiency also to minimum ionising particles are the next efforts needed to qualify purely digital sensors for future applications in scientific experiments and technology.

The very small size of these novel sensors involves an increase of three or more orders of magnitude of the pixel density compared to current state-of-the-art. This implies that conventional readout methods cannot be efficiently used. The challenge of reading out such high pixel densities has not been addressed in this paper; however, we acknowledge that this issue will require a dedicated research effort to devise appropriate strategies and architectures both with a physical layout approach (e.g. stacked readout layers) and application of near detector learning methods (e.g. machine learning techniques).